\documentclass[12pt]{article}

\usepackage[final]{graphicx}
\usepackage{amssymb}
\begin{document}

\author{M. Castelnovo, J.-F. Joanny\\
\textit{Institut Charles Sadron (CNRS UPR $022$), 6 rue
Boussingault,}\\
\textit{\ 67083
Strasbourg Cedex, France.}}
\title{Complexation between oppositely
charged polyelectrolytes: beyond the Random Phase Approximation}

\maketitle

\textbf{PACS Numbers}\\
\hspace*{1cm}82.35.Rs Polyelectrolytes \\
\hspace*{1cm}82.35.Jk Copolymers, phase transitions, structures \\
\hspace*{1cm}64.70.Nd Structural transitions in nanoscale materials.\\
\begin{abstract}
We consider the phase behavior of polymeric systems by calculating the structure factors beyond the Random Phase Approximation. The effect of this correction to the mean-field RPA structure factor is shown to be important in the case of coulombic systems. Two examples are given: simple electrolytes and mixtures of incompatible oppositely charged polyelectrolytes. In this last case,  all former studies predicted an enhancement of compatibility for increasing charge densities; we also describe the complexation transition between the polyelectrolytes. We determine a phase diagram of the polyelctrolyte mixture that includes both complexation and incompatibility.
\end{abstract}

\section{Introduction}
Most polymer blends are incompatible and phase separate in a wide range
of their phase diagram into two phases each containing essentially one
of the components of the mixture. The physical origin of  the segregation 
of the
polymers is their
low mixing entropy of order $1/N$, where $N$ is the number of monomers per
chain, which
cannot overcome any
weak enthalpic interaction. A slight dilution of the two polymers in a solvent
does not change the
qualitative picture of the phase separation,
except that the two phases in coexistence are diluted.
The chemical mismatch between two polymers is measured by 
a Flory-Huggins parameter $\chi_{AB}$ that gives the contribution to the free energy $F_{enthalpic}=kT\, \chi_{AB}\, \phi_A\phi_B$ where the
monomer concentrations
of polymers $A$ and $B$ are
denoted by $\phi_A$ and $\phi_B$. For most polymer pairs, $\chi_{AB}$ is
positive leading to incompatibility. There exists however some
cases where $\chi_{AB}$ is negative and where the polymers are compatible
and form
a single homogeneous phase\cite{kwei}. A way to enhance the compatibility in
solution is
to
add some electrical charges
on the polymers\cite{khokhlov}. The simplest case is a mixture between a
polyelectrolyte and a neutral polymer. In addition
to the ionic charges
along the polyelectrolyte backbone, the solution contains
small counterions that insure the electroneutrality. Their entropy is
sufficient
to
overcome the enthalpic interaction between the backbones since it is
proportional to the
monomer
concentration independent of the polymer 
length: a phase separation is unfavorable
as
the counterions loose too much 
entropy by being confined in one of the phases.
One can also directly take advantage of
the electrostatic interaction by charging
both polymers with ions of opposite charges.
Such experiments have been performed by
Djadoun \textit{et al.}\cite{mora} by mixing
acidic and basic copolymers. If the number of charged groups along
the copolymers is large enough,
a single homogeneous phase is observed.
The corresponding 
ternary phase diagram is sketched
in Figure 1.a. It is qualitatively well described by the theory of Khokhlov
\textit{et al.}\cite{khokhlov} for low charge densities along the polymer.
For higher polymeric charges, the shape of the phase diagram is
completely different, and is sketched in Figure 1.b: the two polymers form
complexes leading to a polymer-solvent type of phase separation. In this case
the Flory parameter is no longer
relevant since the phase behavior is governed by electrostatics. \\
Polyelectrolyte complexes have been studied extensively for thirty years
\cite{complex}. Nevertheless a clear theoretical
description
is not yet available.
A
so-called "host" model has been proposed by 
Nordmeier \textit{et al.}\cite{nordmeier};
the complexation is described as an ion binding between the polymers.  This model is also invoked in
a number of experiments \cite{tsuchida} and should 
be realistic for
highly charged semi-flexible polyelectrolytes. The first attempt to describe 
complexes  between fully
flexible weakly charged polyelectrolytes is due to Borue \textit{et al.}\cite{igor}. In this approach, the complex is
stabilised
by attractive interactions induced by charge
fluctuations. We have used this model recently to describe the complexation
between adjacent layers in polyelectrolyte multilayers \cite{castel}.\\ Our
purpose in this paper is to describe with the same
model the transition from an incompatibility phase diagram (Fig.1a) to a
complexation phase diagram (Fig.1b) when the charge density along the
polymer chains
is increased. The general formalism is introduced in section \ref{genfr}. It relies on a field
theoretical formulation of the Random Phase Approximation. In section
\ref{electrolyte}, we apply the method to a
simple electrolyte as an example of the new results brought by the theory. Then
it is applied to a charged polymer
mixture where the two types of chains
show chemical mismatch. We discuss the calculated
structure factors and 
the phase diagram of the solution in section 
\ref{discussion}. In the last section, we
present some concluding remarks and discuss some possible issues.

\section{General theoretical framework}
\label{genfr} Our aim in this section is to introduce the formalism that we use
throughout
the paper. We introduce a
Hamiltonian $H[\phi(\mathbf{r})]$ that can be written as a functional of the
local field $\phi$ (below $\phi$ is the local concentration or the composition
of the polymer mixture).
The thermodynamic properties of this system are given by the partition function \cite{luben}:
\begin{equation}
\label{part f} Z[h(\mathbf{r})]=\int D \phi (\mathbf{r}) \exp [-\frac{H[\phi (\mathbf{r})]\, -\int d\mathbf{r} h(\mathbf{r})\phi (\mathbf{r})}{kT}]
\end{equation}
where we introduce an external field $h(\mathbf{r})$. For
the rest of this paper we use $kT$ as our energy unit ($kT=1$)
to simplify the notations. The $n$-point correlation functions of the field
are:
\begin{equation}
G(\mathbf{r}_1,\mathbf{r}_2,\dots \mathbf{r}_n)\equiv \langle \phi(\mathbf{r}_1)\phi(\mathbf{r}_2)\dots \phi (\mathbf{r}_n)\rangle=\frac{1}{Z[0]}\frac{\delta ^n Z[h(\mathbf{r})]}{\delta h(\mathbf{r}_1)\, \delta h(\mathbf{r}_2)\dots \delta h(\mathbf{r}_n)} \bigg\arrowvert_{h(\mathbf{r})=0}
\end{equation}
In this
paper we focus on the two-point correlation function (the structure factor).
It can be calculated
as a derivative of the
thermodynamic potential $\Gamma[\bar{\phi}(\mathbf{r})]$ which is the
Legendre transform of the free energy $F[h(\mathbf{r})]=-\log Z[h(\mathbf{r})] $:
\begin{equation}
\Gamma [\bar{\phi} (\mathbf{r}) ] \equiv  F[h(\mathbf{r})]+\, \int d\mathbf{r} h(\mathbf{r})\bar{\phi}(\mathbf{r})
\end{equation}
We call
$\bar{\phi}(\mathbf{r})$ the thermodynamic average of the field $\phi
(\mathbf{r})$.
The correlation function is obtained as:
\begin{equation}
\label{gdef} G^{-1}(\mathbf{r}_1,\mathbf{r}_2)=\frac{\delta ^2\, \Gamma [\bar{\phi}(\mathbf{r})]}{\delta\bar{\phi}(\mathbf{r}_1) \, \delta\bar{\phi}(\mathbf{r}_2)} \bigg\arrowvert_{\bar \phi (\mathbf{r})}
\end{equation}
For any realistic Hamiltonian, the calculation of the path integral involved in
Eq.\ref{part f} cannot be
performed exactly.
The simplest approximation is to compute the partition function at
the saddle point of its integrand: this is the mean field approximation. It
neglects the contribution of all other paths to the integral in the partition
function.
This
approximation leads to rather poor results when correlation effects are
important. One can improve the approximation 
by taking into account the contribution of the fluctuations around the saddle
point path at the gaussian level. This is
the so-called Random Phase Approximation (RPA). We present here a 
standard route
to derive the RPA thermodynamic potential.\\
The field can be decomposed into two parts $\phi (\mathbf{r})= \phi
_{MF}(\mathbf{r}) +\delta \phi (\mathbf{r})$ where the mean field part stems
from the minimisation equation:
\begin{equation}
h(\mathbf{r})=\frac{\delta \, H[\phi (\mathbf{r})]}{\delta \phi (\mathbf{r})}\bigg\arrowvert_{\phi _{MF}}
\end{equation}
Upon expansion of
the Hamiltonian around the saddle point at second order in the fluctuations
of the field, the partition function is rewritten as:
\begin{equation}
Z[h(\mathbf{r})]=Z_{MF}[h(\mathbf{r})]\, \int D\delta \phi (\mathbf{r}) \exp [-\frac{1}{2} \, \int d\mathbf{r}_1 d\mathbf{r}_2\, \delta \phi (\mathbf{r}_1)\, G^{-1}_{RPA}(\mathbf{r}_1 ,\mathbf{r}_2)\, \delta \phi (\mathbf{r}_2)]
\end{equation}
where we introduce the mean field partition function and the mean field
two-point correlation function:
\begin{eqnarray}
Z_{MF}[h(\mathbf{r})] & = & \mathrm{const}. \exp [-H[\phi_{MF} (\mathbf{r})]\, +\int d\mathbf{r} h(\mathbf{r})\phi_{MF} (\mathbf{r})] \nonumber \\
\label{grpa} G^{-1}_{RPA}(\mathbf{r}_1, \mathbf{r}_2) & = & \frac{\delta^2 \, H[\phi (\mathbf{r})]}{\delta \phi (\mathbf{r}_1) \, \delta \phi (\mathbf{r}_2)} \bigg\arrowvert_{\phi_{MF} (\mathbf{r})}
\end{eqnarray}
Note that $G^{-1}_{RPA}$ depends \textit{a priori} on the mean field path $\phi
_{MF}$.
One can then calculate
the RPA thermodynamic potential:
\begin{equation}
\Gamma[\bar{\phi} (\mathbf{r})]=H[\bar{\phi} (\mathbf{r})] -\log \left[ \int D\delta \phi (\mathbf{r})\exp [-\frac{1}{2} \, \int d\mathbf{r}_1 d\mathbf{r}_2\, \delta \phi (\mathbf{r}_1)\, G^{-1}_{RPA}(\mathbf{r}_1 ,\mathbf{r}_2)\, \delta \phi (\mathbf{r}_2)] \right]
\end{equation}
with $\bar\phi (\mathbf{r})=-\frac{\delta F[h(\mathbf{r})]}{\delta h(\mathbf{r})}$.
Using
Eq.\ref{gdef}, the inverse two-point correlation function, or the inverse
structure factor reads:
\begin{eqnarray}
\lefteqn{G^{-1}(\mathbf{r}_1 ,\mathbf{r}_2)=G^{-1}_{RPA}(\mathbf{r}_1 ,\mathbf{r}_2)} \nonumber\\
 & & -\frac{\delta^{2}}{\delta \bar\phi (\mathbf{r}_1)\delta\bar\phi (\mathbf{r}_2)} \left\{\,\log \left[ \int D\delta \phi (\mathbf{r})\exp [-\frac{1}{2} \, \int d\mathbf{r}' d\mathbf{r}'' \, \delta \phi (\mathbf{r}' )\, G^{-1}_{RPA}(\mathbf{r}' ,\mathbf{r}'')\, \delta \phi (\mathbf{r}'')] \right]
\right\}  \label{gdefgen} \nonumber\\
\end{eqnarray}
The RPA formalism is
of common use in polymer physics when one has to deal with rather dense systems where concentration fluctuations are known to be weak\cite{degennes},\cite{leibler}. 
One can calculate both the thermodynamic potential and the structure factors.
The free energy 
is useful to discuss macroscopic phase separation, but the knowledge of the
structure factor is necessary if one wants to study properties at a finite
lenth scale such as a  micro-phase separation.
For this purpose,
several theoretical approaches
only consider the
first term of Eq.\ref{gdefgen}, which corresponds to a mean field approximation for the inverse structure factor: fluctuation effects are neglected.
In
most cases,
this is sufficient to discuss the 
phase separations of interest.\\
However for some coulombic systems, correlation effects might be relevant.
Consider for example the case of a simple monovalent electrolyte. We
show in the next section that neglecting fluctuations in the RPA calculation of
the structure factor, which is the spirit of the Debye-H\"{u}ckel theory of electrolytes (DH), leads to a constant density-density structure factor that does not satisfy all the sum rules imposed by the electroneutrality. This observation is at the origin of the Generalized Debye-H\"{u}ckel (GDH) theory of Lee and Fisher which produces a $q$-dependent structure factor \cite{lee}.\\
One can qualitatively understand why a mean field approximation misses some
relevant properties in coulombic system: because of the electroneutrality condition,
no
pure coulombic term can appear in
a mean field thermodynamic potential in the absence of
external field, since fluctuations around the electroneutral state are not included.\\
The RPA calculation of the thermodynamic potential is also called the one loop
approximation, since it corresponds to a resumation of all the one loop diagrams of the  associated field theory for this potential. In the following, we
call the result of Eq.\ref{gdefgen} the one loop approximation (OLA) of the inverse structure factor, to distinguish it from the
standard RPA approximation,
that includes only the zeroth order term in a loop expansion.
\section{Structure
factor of a simple electrolyte}
\label{electrolyte}We consider in this section a simple monovalent electrolyte solution in water composed
of discrete point-like ions embedded in a continuous dielectric background of permittivity $\epsilon$. For
a sake of simplicity, we neglect the effect of the hard core potential since
we only
want to show the effect of the concentration fluctuations
and not to discuss quantitatively
the phase diagram. Simple electrolytes have been
investigated quite recently by Netz \textit{et al.} \cite{netz} and Frusawa
\textit{et al.} \cite{frusawa} independently using field theoretical methods.
The
N-particle problem has been 
mapped exactly onto a field theory with two fields which correspond to collective variables: the first one is a density field and the
second 
one is an auxiliary field which acts as a complex potential on the ions. At the saddle point of the auxiliary
field, the partition function is written $Z=\int Dn_+ (\mathbf{r}) Dn_- (\mathbf{r})\exp [-H[n_+ (\mathbf{r}),n_- (\mathbf{r})]]$ with the Hamiltonian:
\begin{eqnarray}
H[n_+ (\mathbf{r}),n_- (\mathbf{r})] & = & \int d\mathbf{r} \bigg\{ n_+ (\mathbf{r}) (\log n_+ (\mathbf{r}) -1)
\,
 +n_- (\mathbf{r}) (\log n_- (\mathbf{r})-1) \, 
\nonumber\\
& & +\frac{l_B}{2}\int d\mathbf{r}' \frac{(n_+ (\mathbf{r})-n_- (\mathbf{r}))(n_+ (\mathbf{r}')-n_- (\mathbf{r}'))}{|
\mathbf{r}-\mathbf{r}'| }
\bigg\} \nonumber\\
\end{eqnarray}
The
Bjerrum length is defined here by
$l_B =\frac{e^2 }{4\pi \epsilon}$; it
measures the strength of electrostatic interactions. This hamiltonian can be
used as 
the starting point of a density functional theory where the equilibrium density
fields
result from a balance between  entropy and electrostatics. At
this level, we neglect the fluctuations of the auxiliary field and investigate
the effect of composition fluctuations. We show below that
we recover this way the
results of the Generalized Debye-H\"{u}ckel theory if we include the
composition fluctuations at the level of a one loop approximation,
and the classical results of the Debye-H\"{u}ckel theory if we neglect
composition fluctuations.
Let $n_{\sigma}(\mathbf{r})=\bar{n}_{\sigma}(\mathbf{r})+\delta n(\mathbf{r})$ with $\sigma =+,-$. Then using the definition of section \ref{genfr}, the partition function is:
\begin{equation}
{Z=Z_{MF}\, \int D\delta n_+ (\mathbf{r}) D\delta n_- (\mathbf{r})\, \exp
[-\Delta H[\bar{n}_+ ,\bar{n}_- ,\delta n_+ ,\delta n_- ]]}   \nonumber 
\end{equation}
where the effective hamiltonian is defined as
\begin{equation}
\Delta H[\bar{n}_+ ,\bar{n}_- ,\delta n_+ ,\delta n_- ]=\frac{1}{2}
\sum_{\sigma ,\sigma '}\int d\mathbf{r} d\mathbf{r}' \, \delta
n_{\sigma} (\mathbf{r})G^{-1}_{RPA\, \, \sigma ,\sigma '}(\mathbf{r},\mathbf{r}')\delta n_{\sigma '} (\mathbf{r}')] \nonumber\\ 
\end{equation}
The RPA inverse structure factor matrix is given by:
\begin{equation} 
G^{-1}_{RPA\, \, \sigma ,\sigma '}(\mathbf{r},\mathbf{r}')=\left(\begin{array}{cc}\frac{\delta (\mathbf{r}-\mathbf{r}')}{\bar{n}_+ (\mathbf{r})} +\frac{4\pi l_B}{| \mathbf{r}-\mathbf{r}' |}& -\frac{4\pi l_B}{| \mathbf{r}-\mathbf{r}' |} \\
 -\frac{4\pi l_B}{| \mathbf{r}-\mathbf{r}' |} & \frac{\delta (\mathbf{r}-\mathbf{r}')}{\bar{n}_- (\mathbf{r})}+\frac{4\pi l_B}{| \mathbf{r}-\mathbf{r}' |}
\end{array}\right)
\end{equation}
One can easily generalize Eq.\ref{gdefgen} to a multi-component system to
obtain the inverse structure factor matrix in the one loop approximation:
\begin{eqnarray}
G^{-1}_{\sigma ,\sigma '}(\mathbf{r}, \mathbf{r}') & = & G^{-1}_{RPA\, \, \sigma ,\sigma '} (\mathbf{r}, \mathbf{r}')\nonumber\\
 & & +\bigg\langle \frac{\delta ^2\, \Delta H}{\delta \bar{n}_{\sigma}(\mathbf{r}) \delta \bar{n}_{\sigma '}(\mathbf{r}')} \bigg\rangle
 +\bigg\langle \frac{\delta \Delta H}{\delta \bar{n}_{\sigma}(\mathbf{r})}\bigg\rangle \bigg\langle \frac{\delta \Delta H}{\delta \bar{n}_{\sigma '}(\mathbf{r}')}\bigg\rangle -\bigg\langle \frac{\delta \Delta H}{\delta \bar{n}_{\sigma}(\mathbf{r})} \frac{\delta \Delta H}{\delta \bar{n}_{\sigma '}(\mathbf{r}')}\bigg\rangle 
\end{eqnarray}
where the average is performed with the gaussian weight $\exp [-\Delta
H[\bar{n}_+ ,\bar{n}_- ,\delta n_+ ,\delta n_- ]]$ of the fluctuating variables $\delta n_+ ,\delta n_-
$. The
average densities $\bar{n}_+ ,\bar{n}_- $ are
constant in
a homogeneous phase. Note that  the densities must be considered as constant
only at the
step of averages, after the formal derivations with respect to inhomogeneous
densities; otherwise the contributions of the fluctuations vanishes.\\
It is convenient to study
the Fourier transform of the structure factors.
Each element of the matrix is then rewritten as:
\begin{eqnarray}
\lefteqn{G^{-1}_{\sigma ,\sigma '}(\mathbf{q})=G^{-1}_{RPA \, \, \sigma ,\sigma '}(\mathbf{q})} \nonumber\\
 & & -\frac{1}{2\, \bar{n}^{2}_{\sigma} \bar{n}^{2}_{\sigma '}}\int \frac{d^3 \mathbf{q}_1}{(2\pi)^3 }\, G_{RPA \, \, \sigma,\sigma '}(\mathbf{q}_1) G_{RPA \, \, \sigma,\sigma '}(\mathbf{q}-\mathbf{q}_1)\, +\frac{\delta_{\sigma ,\sigma '}}{\bar{n}_{\sigma}}\int \frac{d^3 \mathbf{q}_1}{(2\pi)^3 }\, G_{RPA \, \, \sigma,\sigma '}(\mathbf{q}_1) \label{gq}\nonumber\\
\end{eqnarray}
where the gaussian average has been already performed. All the integrals involved in Eq.\ref{gq} can be calculated analytically by using Feynman integral
techniques.
The diagonal terms are diverging
at high $q$. This divergence can be removed by substracting the contribution of
the neutral system (perfect gas). The structure factor matrix
of the electrolyte solution reads:
\begin{equation}
G^{-1}_{\sigma ,\sigma '}(\mathbf{q})=\left( 
\begin{array}{cc}
\frac{1}{\bar{n}}+\frac{4\pi l_{B}}{\mathbf{q}^2 }-\Delta G^{-1}(\mathbf{q}) & -\frac{4\pi l_{B}}{\mathbf{q}^2 }-\Delta G^{-1}(\mathbf{q})\\
-\frac{4\pi l_{B}}{\mathbf{q}^2 }-\Delta G^{-1}(\mathbf{q}) & \frac{1}{\bar{n}}+\frac{4\pi l_{B}}{\mathbf{q}^2 }-\Delta G^{-1}(\mathbf{q})
\end{array}\right)
\end{equation}
The electroneutrality condition imposes the equality of homogeneous densities $\bar{n}_+ =\bar{n}_- =\bar{n}$. The correction term is then given by:
\begin{equation}
\Delta G^{-1}(\mathbf{q})=\frac{2\pi l_B ^2}{q}\, \arctan \left(\frac{q}{2\kappa}\right)
\end{equation}
with the Debye-H\"{u}ckel length defined by $\kappa ^2=8\pi l_B \bar{n}$.
The effect
of the fluctuation contribution
can be drawn from the density-density structure factor that is
calculated from
the partial structure factors $G_{\bar{n}_{tot},\, \bar{n}_{tot}}=G_{\bar{n}_+
,\, \bar{n}_+ }+G_{\bar{n}_- ,\, \bar{n}_- }+2G_{\bar{n}_+ ,\, \bar{n}_-
}$.
\begin{equation}
\label{densdens}G_{\bar{n}_{tot},\, \bar{n}_{tot}}(\mathbf{q})=\frac{2\bar{n}}{1-2\bar{n}\Delta G^{-1}(q)}
\end{equation}
If we neglect 
the fluctuations at all orders,the density-density structure
factor is constant.
The fluctuations induce
a decrease
of the structure factor for increasing wave vectors. Notice that the value
value at $q=0$ is consistent with the Debye-H\"{u}ckel
polarization energy $F_{pol}=-\frac{\kappa^3}{12\pi}$. This is expected
since the results of the OLA and the RPA are the same for the free energy. \\
This natural generalization of the Debye-H\"uckel theory
to structure factors must be compared to the Generalised Debye-H\"uckel theory
proposed by Lee and Fisher
(\cite{lee}. Starting from the same constatation on the 
behaviour of the
density-density structure factor in the Debye-H\"uckel
theory, they solve the Debye-H\"uckel
equation for the electrostatic potential with
modulated densities. When the amplitude of the modulation vanishes at the end
of the calculation, they obtain a $q$-dependent structure factor. Notice that
they take into account the
finite radius of the
ions that we neglect here.
Nevertheless their
procedure is
very similar to ours.
In fact, the correlation length that we can define from the density-density structure factor is exactly the same in the limit of vanishing ion radius. More precisely the $q=0$ susceptibility and the correlation length are in the limit of vanishing ion
radius:
\begin{eqnarray}
\label{chi}
\chi ^{-1}(0) = 1-\frac{\kappa l_B }{4},\quad  \xi ^2 =
\frac{\chi (0)l_B }{48\kappa} 
\end{eqnarray}
in agreement with the
results of Lee and Fisher. It is clear from Eqs.\ref{densdens} and \ref{chi} that the
"One loop approximation"
is valid as long as $\frac{\kappa l_B}{4}<1$. If it is not the case, one has to
consider
higher orders in the loop expansion. The perturbation
parameter of the theory is $\zeta=\frac{\kappa l_B}{4}$.\\ 
We do not discuss further the properties of the structure factor because it
goes beyond the scope of this paper. We note
however that simple tests do not reveal any inconsistencies, in particular with the  Stillinger-Lovett sum rules \cite{stil}.
\section{Complexation and incompatibility in mixtures of oppositely charged polyelectrolytes}
\subsection{Inverse structure factor matrix}
\label{complex}
In this section we generalise the results obtained in the previous section
for simple electrolytes to polyelectrolyte complexes which are mixtures of
polyelectrolytes of opposite charges.
We use the structure factors to study the stability of the polyelectrolyte
mixture and compare our results to the experimental work 
of Djadoun \textit{et al.}\cite{mora} mentioned in the introduction. In this
work, the charge of acidic and basic copolymers is progressivley increased.
The neutral
backbones of those copolymers exhibit a strong incompatibility. Starting from
an incompatibility phase diagram at zero charge, the compatibility range (solubilisation)
increases with the charge and
finally turns into
a complexation phase diagram where the two copolymers are in the same neutral
dense phase in equilibrium with almost pure water because
of the strong electrostatic attraction between them.\\
In order to describe this
system,
we consider a mixture 
of two polyelectrolytes
A and B having the same linear charge density $f$ but of opposite charges
\cite{barrat}. The charge density is smeared out along the
backbone and the solvent is modelled
as a continuous dielectric background of permittivity $\epsilon$. A
concentration $n_+ +n_- =2n$ of small point-like ions of opposite charges is added to the solution.
We study only a symmetric system and the
monomer
concentrations are $c_+ =c_- =c$. Each chain is composed of $N$ monomers of
size $a$. If
the electrostatic interactions are switched off,
the short range interactions between polymers are characterised
by three Flory parameters $\chi_{AS},\chi_{BS}$ and $\chi_{AB}$. For a sake of
simplicity, we will only consider both polymers are in a $\theta$ solvent and
fix the values
$\chi_{AS}=\chi_{BS}=\frac{1}{2}$. The
last parameter $\chi_{AB}\equiv
\chi$ characterises the chemical mismatch between the two polymers and is
varied
independently.

The
total Hamiltonian of the solution can then be written as:
\begin{eqnarray}
H[c_+ (\mathbf{r}),c_- (\mathbf{r}),n_+ (\mathbf{r}),n_- (\mathbf{r})] & = & \int d\mathbf{r} \Bigg\{
n_+ (\mathbf{r}) (\log n_+ (\mathbf{r})-1)+n_- (\mathbf{r}) (\log n_- (\mathbf{r})-1)\nonumber\\
& & +\frac{a^2}{24}\left(\frac{|\nabla c_+ (\mathbf{r})|^2}{c_+ (\mathbf{r})}+\frac{|\nabla c_- (\mathbf{r})|^2}{c_- (\mathbf{r})}\right) \nonumber\\
 & & \, \hspace{2cm}+\frac{l_B}{2}\int d\mathbf{r}' \frac{\rho_z (\mathbf{r})\, \rho_z (\mathbf{r}')}{|\mathbf{r}-\mathbf{r}'|} \nonumber\\
 & & \, \hspace{3.5cm}+\chi c_+ (\mathbf{r}) c_- (\mathbf{r})\label{hamil}\Bigg\}
\end{eqnarray}
The
charge density is defined here as
$\rho_z (\mathbf{r})=f(c_+ (\mathbf{r})-c_- (\mathbf{r}))+n_+ (\mathbf{r})-n_- (\mathbf{r})$.
 The first term of Eq.\ref{hamil} is the
translational entropy of the small
ions. The second term is
the conformational entropy of the polymer chains
in the limit of infinite chain length. In this limit the
conformational entropy can be calculated from a
ground state dominance approximation and it is given by the
Lifshitz formula
\cite{desclois}. In the limit of infinite chain length, the
translational entropy of the polymer chains
is negligibly small.
The third term is the electrostatic interaction between the charged components of the
solution.
Finally the last term models the enthalpic interactions that give rise to the
incompatibility between the polymers.

From
this model Hamiltonian we want to 
calculate
various structure factors in order to discuss the stability limits of the
solution and its
phase diagram. As for simple electrolytes, the calculation is performed
at the level of the one loop approximation  presented in
section \ref{genfr}. \\
For a
sake of simplicity, we include the enthalpic interaction between the polymers
only at the mean field level, or equivalently at the RPA level in the
calculation of 
the structure factor. We believe that this approach gives the correct
physical
result because we expect fluctuations to be dominated by the connectivity of the chains. This calculation can also be used as a
starting point to
study various systems involving
gaussian polymer chains
in external
potentials (in order to take into account excluded
volume or solvent effects for example). Note that the contribution of
the Flory parameter
to the 
fluctuation can also be computed
but the calculations are lengthy and much more
tedious.\\
We first ignore the short range interactions, $\chi=0$.
In order to find the fluctuation contributions to the structure
factor, let $c_{\sigma} (\mathbf{r})=\bar{c}_{\sigma} (\mathbf{r})+\delta
c_{\sigma} (\mathbf{r})$ and $n_{\sigma} (\mathbf{r})=\bar{n}_{\sigma}
(\mathbf{r})+\delta n_{\sigma} (\mathbf{r})$. By expanding the Hamiltonian up to second order in the composition fluctuations, we calculate the thermodynamic potential and the resulting inverse structure factor. Within the one loop approximation, the inverse structure factor
reads:
\begin{eqnarray}
\label{ola}\lefteqn{G^{-1}_{ \rho, \, \, \sigma ,\sigma '}(\mathbf{r}, \mathbf{r}')=G^{-1}_{RPA\, \rho  ,\, \, \sigma ,\sigma '} (\mathbf{r}, \mathbf{r}')}\nonumber\\
 & & +\bigg\langle \frac{\delta ^2\, \Delta H}{\delta \bar{\rho}_{\sigma}(\mathbf{r}) \delta \bar{\rho}_{\sigma '}(\mathbf{r}')} \bigg\rangle
 +\bigg\langle \frac{\delta \Delta H}{\delta \bar{\rho}_{\sigma}(\mathbf{r})}\bigg\rangle \bigg\langle \frac{\delta \Delta H}{\delta \bar{\rho}_{\sigma '}(\mathbf{r}')}\bigg\rangle -\bigg\langle \frac{\delta \Delta H}{\delta \bar{\rho}_{\sigma}(\mathbf{r})} \frac{\delta \Delta H}{\delta \bar{\rho}_{\sigma '}(\mathbf{r}')}\bigg\rangle 
\end{eqnarray}
with the short-hand notation $\bar{\rho}_{\sigma}=(\bar{c}_+ ,\bar{c}_- ,\bar{n}_+ ,\bar{n}_- )$. The quantity $\Delta H$ is the second order term in the expansion of the Hamiltonian $H$. As in the previous section, the average is performed with the gaussian weight $\exp [-\Delta H]$ of the fluctuating
densities. The average densities $\bar{\rho}$ can in the calculation of the
averages be set equal to their uniform values in the solution,
 $c$ and $n$ respectively for the monomers and the small ions.
 The RPA inverse structure factor has been already calculated by several
authors \cite{igor}, it is given in reciprocal space by:
\begin{equation}
\label{grpa2}G^{-1}_{RPA \, \rho, \, \, \sigma, \sigma '}=\left(\begin{array}{cccc}
\frac{q^2 a^2}{12\bar{c}}+\frac{4\pi l_B f^2}{q^2} & -\frac{4\pi l_B f^2}{q^2} & \frac{4\pi l_B f}{q^2} & -\frac{4\pi l_B f}{q^2}\\
-\frac{4\pi l_B f^2}{q^2} & \frac{q^2 a^2}{12\bar{c}}+\frac{4\pi l_B f^2}{q^2} & -\frac{4\pi l_B f}{q^2} & \frac{4\pi l_B f}{q^2}\\
\frac{4\pi l_B f}{q^2} & -\frac{4\pi l_B f}{q^2} & \frac{1}{\bar{n}}+\frac{4\pi l_B}{q^2} & -\frac{4\pi l_B}{q^2}\\
-\frac{4\pi l_B f}{q^2} & \frac{4\pi l_B f}{q^2} &  -\frac{4\pi l_B}{q^2} & \frac{1}{\bar{n}}+\frac{4\pi l_B}{q^2}
\end{array}\right)
\end{equation}
 We give below the values of the corrections to the RPA in Eq.\ref{ola}
for the various components of the inverse structure factor matrix.\\
The monomer-monomer components read:
\begin{eqnarray}
\label{mg}\lefteqn{\Delta G^{-1}_{c\,\sigma,\sigma '}=}\nonumber\\
& & \frac{a^4}{576}\Bigg\{
-\frac{2}{c^2 _{\sigma}c^2 _{\sigma '}}\int\frac{d^3 \mathbf{q}_1}{(2\pi)^3}\left(\mathbf{q}_1.(\mathbf{q}-\mathbf{q}_1)-\mathbf{q}^2\right)^2
G_{RPA\, c\, \, \sigma,\sigma '}(\mathbf{q}_1)G_{RPA\, c\, \, \sigma,\sigma '}(\mathbf{q}-\mathbf{q}_1)\nonumber\\
& & \, \hspace{1cm}+\frac{48\,\delta_{\sigma,\sigma'}}{c^3 _{\sigma}a^2}\int\frac{d^3 \mathbf{q}_1}{(2\pi)^3}\left(\mathbf{q}^2+\mathbf{q}^2 _1\right)G_{RPA\, c\, \, \sigma,\sigma '}(\mathbf{q}_1)
\Bigg\}
\end{eqnarray}
The correction term to the RPA for the salt-salt components of the inverse
structure factor matrix are given by the same formula Eq.\ref{gq} as for
a simple electrolyte:
\begin{eqnarray}
\label{sg}\lefteqn{\Delta G^{-1}_{n\,\sigma,\sigma '}=}\nonumber\\
 & & -\frac{1}{2\, \bar{n}^{2}_{\sigma} \bar{n}^{2}_{\sigma '}}\int \frac{d^3 \mathbf{q}_1}{(2\pi)^3 }\, G_{RPA\, n\, \sigma,\sigma '}(\mathbf{q}_1) G_{RPA \, n\, \sigma,\sigma '}(\mathbf{q}-\mathbf{q}_1)\, +\frac{\delta_{\sigma ,\sigma '}}{\bar{n}_{\sigma}^{3}}\int \frac{d^3 \mathbf{q}_1}{(2\pi)^3 }\, G_{RPA \, n\, \sigma,\sigma '}(\mathbf{q}_1) \nonumber\\
\end{eqnarray}
Note however that 
the RPA structure factor is different from the simple electrolyte case. \\
Finally the corrections to the crossed
salt-monomer terms of the inverse structure factor matrix  are given by:
\begin{equation}
\label{msg}\Delta G^{-1}_{c_{\sigma},n_{\sigma '}}=\frac{a^2}{24c_{\sigma}^2 n_{\sigma'}^2 }\int \frac{d^3 \mathbf{q}_1}{(2\pi)^3 }\left(\mathbf{q}_1.(\mathbf{q}-\mathbf{q}_1)-\mathbf{q}^2\right)G_{RPA \,{c_{\sigma},n_{\sigma'}}}(\mathbf{q}_1)G_{RPA \,{c_{\sigma},n_{\sigma'}}}(\mathbf{q}-\mathbf{q}_1)
\end{equation}
All the integrals involved in Eqs.\ref{mg}, \ref{sg} and \ref{msg} can be calculated using Feynman integrals
techniques.
The analytical expressions
of these integrals are reported in appendix A. 

We can therefrom calculate the
inverse structure factor at the level of the one loop approximation.
As the inverse structure factor matrix is symmetric, we
only give the independent terms:
\begin{eqnarray}
G^{-1}_{OLA\,c_+ ,c_+  }=G^{-1}_{OLA\,c_- ,c_-  }& = & \frac{q^2 a^2}{12c}+\frac{4\pi l_B f^2}{q^2}+\frac{f^2}{c}\alpha (q)-f^4\beta (q)\\
G^{-1}_{OLA\,c_+ ,c_- }& = & -\frac{4\pi l_B f^2}{q^2}-f^4\beta (q)\\
G^{-1}_{OLA\,n_+ ,n_+ }=G^{-1}_{OLA\,n_- ,n_- } & = & \frac{1}{n}+\frac{4\pi l_B}{q^2}-\delta (q)\\
G^{-1}_{OLA\,n_+ ,n_- }& = & -\frac{4\pi l_B}{q^2}-\delta (q)\\
G^{-1}_{OLA\,c_+ ,n_+ }=G^{-1}_{OLA\,c_- ,n_- } & = & \frac{4\pi l_B f}{q^2} -f^2\gamma (q)\\
G^{-1}_{OLA\,c_+ ,n_- }=G^{-1}_{OLA\,c_- ,n_+ } & = & -\frac{4\pi l_B f}{q^2} -f^2\gamma (q)
\end{eqnarray}
The quantities $\alpha,\beta,\gamma,\delta$ are related to the Feynman
integrals 
of appendix A by:
\begin{eqnarray}
\alpha(q) & = & \frac{24\pi l_B}{a^2}I_2 (q)\nonumber\\
\beta(q) & = & \frac{1}{2}\left(\frac{48\pi l_B}{a^2}\right)^2 I_1 (q)\\
\gamma(q) & = &\frac{96(\pi l_B )^2}{a^2} I_4 (q)\nonumber\\
\delta(q) & = & 8(\pi l_B)^2 I_3 (q)\nonumber
\end{eqnarray}
\subsection{Phase diagram and discussion of the experimental results}
\label{discussion}We have obtained in the preceding section the inverse structure factor matrix in the particular basis $\{c+,c_-,n_+,n_-\}$. In order to discuss the phase diagram of the system, it is more convenient to calculate the structure factor matrix for the  variables $\{c_{tot},c_z,n_{tot},n_z\}$ denoting respectively the  total monomer density ($c_{tot}=c_+ +c_-$), the  total monomer charge density ($c_z =f(c_+-c_-)$), the  total small ion density ($n_{tot}=n_++n_-$) and the total small ion charge density ($n_z=n_+-n_-$). Indeed the density-density structure factor is essential to predict a phase separation such a complexation which is a polymer-solvent phase separation; it is associated to  a singular behavior of this structure factor. On the other hand, the  incompatibility between the chains, is a polymer-polymer phase separation, and is associated to  a singularity in the charge-charge structure factor, since the polymer charge is the only way to distinguish the two types of chains within our model. Previous studies have focused on this last structure factor because it exhibited  a singularity while the density-density structure factor was regular. We show below that the effect of the composition fluctuations is to induce an attraction between the chains that counterbalances in some parameter range the chemical mismatch between the chains. This effect was ignored in all the previous studies that stay at the RPA or mean field level for the structure factor.

The change of variables and the inversion of the matrix $G^{-1}_{OLA}(q)$ can be performed analytically. At this point we add at the RPA level the Flory parameter $\chi$. The structure factor bloc matrix for the polymer variables then reads:
\begin{eqnarray}
\lefteqn{G_{OLA\, c_{\sigma},c_{\sigma '}}(q)=}\nonumber\\
& &\left(\begin{array}{cc}
\frac{1}{\frac{1}{2c}\left(\frac{q^2 a^2}{12}+f^2\alpha(q)-2f^4c\beta(q)\right)+\frac{\chi}{2}} & 0 \\
0 & \frac{1}{\frac{1}{2f^2c}\left(\frac{q^2 a^2}{12}+f^2\alpha(q)\right)+\frac{4\pi l_B}{q^2+\kappa^2}-\frac{\chi}{2f^2} }
\end{array}\right)\end{eqnarray}
where $\sigma=(tot,z)$.
In this last equation, we have neglected higher order contributions obtained in  the matrix inversion involving the square of correction terms since these orders in the loop expansion have been neglected from the very beginning.\\
If we neglect $\alpha (q)$ and $\beta (q)$ in the structure factor, we recover the results of Khokhlov \textit{et al}\cite{khokhlov}: their discussion is limited to the charge-charge structure factor. We present here a similar analysis by omitting first the effect of the density-density correlations on the phase diagram.  The phase diagram can be drawn in a very transparent and elegant way by using the dimensionless variables introduced first by Borue \textit{et al}\cite{igor}. So far, we have introduced two characteristic lengths:  the Debye-H\"{u}ckel screening length $\kappa^{-1}$ and the correlation length of a dense polyelectrolyte mixture $q^{-1}_*$ ($q^4_*=\frac{96\pi l_B f^2 c}{a^2}$). The first length is associated to the screening of electrostatic interactions by the small ions and the second length is associated to the screening of the same interactions by the polymers. The enthalpic interactions between the neutral backbones of the chains are measured by  the Flory parameter $\chi$ to which we can associate the mean field length $\xi_{\chi}$ defined by \cite{degennes}:
\begin{equation}
\xi^{-2}_{\chi}=\frac{12\chi c}{a^2}
\end{equation}
The physical behaviour of the system is then governed by the following dimensionless ratios:\newpage
\begin{eqnarray}
s & = & \frac{\kappa^2}{q^2_*}\nonumber\\
t & = & \frac{\xi^{-2}_{\chi}}{q^2_*}
\end{eqnarray}
The choice of $q^{-1}_*$ as the unit length is motivated by the fact that we are expecting the connectivities  of the chains to play a dominant role. \\ The stability of the system against the polymer-polymer phase separation is ensured if $1/G_{c_z c_z}(q)$ is positive for all values of $q$. The spinodal line giving the stability limit is obtained when this condition is violated. If the first mode for which $1/G_{c_z c_z}(q)$ becomes negative is $q_0\neq 0$, the solution shows a microphase separation. On the contrary, for $q_0 =0$ there is macrophase separation. The partial phase diagram is shown in Figure 2. It does not take into account yet instabilities of the density-density structure factor.  An analysis based on the stability of the homogeneous phase can only give the equation of the spinodal lines but it cannot be used to predict the precise morphology of the different phases after the transition. A complete phase diagram can only be 
obtained by calculating the free energy of the various possible phases and by looking for the phases that minimise it. In this paper we study only the stability limits  of the homogeneous phase.\\
We checked by numerical evaluation of the structure factor with the exact formula that the correction term $\alpha (q)$ can be neglected. If $s>1$ the system is homogeneous for $st<1$, otherwise it is macroscopically phase separated: chains of opposite charge are segregated into two dilute phases. This is due to the incompatibility of the polymers. If $s<1$ it is possible to have microphase separation. The spinodal line for this transition is given by $t=2-s$. The wave vector at the transition is $q^2_0=q^2_*\frac{t-s}{2}$. It is related to the wavelength of modulation of the mesophase and it tends to infinity as we approach the macroscopic spinodal line $st=1$. Notice that in the limit of high ionic strength ($s\gg 1$) which is the relevant case for mÁost experiments, no microphase separation is expected.

The full phase diagram includes possible instabilities associated to the density-density structure factor. The numerical analysis of this structure factor shows that the correction term $\alpha (q)$ can again be neglected and that the term $\beta (q)$ can be approximated by a constant $\beta (q)\approx \beta (q=0)$. The result of this analysis is that the composition fluctuations induce a macroscopic complexation between the oppositely charged polyelectrolytes when:
\begin{equation}
t<\frac{u}{\,\,(s+2)^{3/2}}
\end{equation}
where we introduce the following dimensionless ratios:
\begin{eqnarray}
u & = & \frac{\xi_{mesh}}{q^{-1}_*}\\
\xi_{mesh} & = & \frac{3}{8\pi a^2c} 
\end{eqnarray}
The length $\xi_{mesh}$ is the mesh size of a gaussian transient network (gaussian chains in semi-dilute regime). This length is relevant to discuss the validity limits of the OLA or the RPA calculations. The main assumption of those approximations is that the chains are gaussian at all length scales. For simple polyelectrolyte solutions, this remains true as long as the electrostatic blob size \footnote{The electrostatic blob size is the length scale at which the typical electrostatic energy of a subunit composed of $\left(\frac{\xi_{el}}{a}\right)^2$ becomes larger than the thermal energy. More precisely $kT\approx \frac{kT\,l_B\,f^2}{\xi_{el}}\left(\frac{\xi_{el}}{a}\right)^4$} is larger than the mesh size of the semi-dilute solution because the electrostatic interactions are screened by the transient network at large length scales\cite{barrat}. Up to irrelevant prefactors this condition is equivalent to $u<1$. In terms of polymer concentration this threshold is written $c\,a^3 \approx \frac{l_B^{4/3}f^{2/3}}{a^{4/3}}$. Therefore the approximation used in this paper is valid at moderately high concentration.\\
The phase diagram of the system is presented in Figure 3. Note that there is no microphase separation associated to the polyelectrolyte complexation. Although the variables used to describe the phase diagram look very  physical, they are not well suited to describe the different phases in terms of concentration. In Figure 4 we translate the $t-s$ diagram into a $n-c$ diagram (all the other parameters being kept constant). The equations of the various spinodal lines are reported in Appendix B.\\
If we now come back to the experiments of Djadoun \textit{et al.} discussed above, our model is able to describe qualitatively the transition from incompatibility to compatibility and finally complexation for increasing polymeric charges. Indeed the parameters $t$ and $s$ scale like:
\begin{eqnarray}
t & \approx & \frac{\chi c^{1/2}}{l_B^{1/2}af}\\
s & \approx & \frac{l_B^{1/2}na}{c^{1/2}f}
\end{eqnarray}
so that the two parameters have the same dependence in the polymer charge density, except for prefactors. If we increase $f$, we move in the phase diagram on a line represented in Figure 3. Starting at low $f$, the solution is in the incompatibility range; increasing $f$ pushes it into the compatibility range and finally complexation takes place. A very rough model of the complex can be made by describing it as a collapsed polymer \cite{castel}. This argument also shows that incompatibility becomes an irrelevant parameter once the complex is formed.

In real systems, the polymers are not symmetric in length and charge, but we believe that this will not affect significantly the general shape of our proposed phase diagram: the neutral dense phases should be similar in composition but we expect the supernatant to be composed of charged soluble species. The description of the supernatant will be the topic of a forthcoming paper.
\section{Concluding remarks}
We have proposed in this article description of polymer mixtures in a solvent that goes beyond the classical Random Phase Approximation and that allows to describe simultaneously in a certain range of parameters both the incompatibility between the two polymers and the formation of polyelectrolyte complexes when the polymers are charged. This one loop approximation has been used to calculate the structure factors. The Random Phase Approximation is commonly used  to discuss phase stability in polymer systems but in the calculation of the structure factors, it misses important correlation effects. As pointed out by Haronska \textit{et al.}\cite{haronska}, the RPA is able to predict only repulsive electrostatic contributions to the mean field effective potential between polymers. Including fluctuations at the order of one loop allows for attractive effective interactions. Using a One Loop Approximation at the level of the structure factors, we were able to include the effect of the fluctuations for two systems: simple electrolytes and mixtures of incompatible oppositely charged polyelectrolytes. For simple electrolytes, we obtain a natural generalization of the Debye-H\"{u}ckel theory which leads to results similar to the one proposed by Lee and Fisher. For polyelectrolytes blends, we were able to describe within a single model the experiments showing a transition from an incompatibility phase diagram to a complexation phase diagram. The transition towards polyelectrolyte complexation is predicted to take place in the range of validity of the OLA, namely in the limit of weakly charged chains\\
The main limitation of this work is that for high enough charge densities the one loop approximation breaks down: the system becomes strongly fluctuating and one should use another formalism. A model based on the formation of ion pairs seems appropriate in this case; whe hope to be able to set up such a model in a future publication.

{\bf Acknowledgments:} We are grateful for discussions to I. Erukhimovitch (Moscow state University). This research was supported by the Deutsche Forschung Gemeinschaft through the Schwerpunkt program "Polyelektrolyte".
\section*{Appendix A}
\subsection*{A.1\ \ Exact expression for the monomer-monomer correction terms}
By replacing the expression of the RPA structure factor into Eq.\ref{mg}, we obtain for the off-diagonal term:
\begin{equation}
\Delta G^{-1}_{c\,+,-}=-\frac{1}{2}\left(\frac{48\pi l_B f^2}{a^2}\right)^2\,I_1 (\mathbf{q})\label{dgpm}
\end{equation}
$I_1 (\mathbf{q})$ is a Feynamn-like integral defined by:
\begin{eqnarray}
\lefteqn{I_1 (\mathbf{q})=}\nonumber\\
& & \int\frac{d^3 \mathbf{q}_1}{(2\pi)^3}\, \frac{ \left( \mathbf{q}_1.(\mathbf{q}-\mathbf{q}_1)-\mathbf{q}^2 \right)^2 }{\mathbf{q}^2 _1 (\mathbf{q}^2 -\mathbf{q}^2 _1)}\frac{1}{[\mathbf{q}^2 _1 (\mathbf{q}^2 _1 +\kappa^2 )+q^4 _* ][(\mathbf{q}-\mathbf{q}_1)^2 ((\mathbf{q}-\mathbf{q}_1 )^2 +\kappa^2 )+q^4 _* ]}\label{i1}
\end{eqnarray}
with the characteristic length $q^{-1}_* $ defined by $q^4 _* =\frac{96\pi l_B f^2 c}{a^2}$. It can be interpreted as the correlation length of a dense mixture of oppositely charged polyelectrolytes in the absence of small ions \cite{castel}.
By reducing the integrand of Eq.\ref{i1} to simple fractions we can expressed the whole integral as a sum of Feynman diagrams. We give here only the expansion at low $q$ because the discussion on the possibility of macro- \textit{or} microphase separation is relevant only in this range, while the exact formula is rather lengthy:
\begin{equation} 
I_1 (q)=\frac{1}{8\pi q_+ q_- (q_+ +q_- )^3}+q^2\,\frac{15q^4_+ +75q^3_+ q_- +116q^2 _+ q^2 _- +75q_+ q^3 _- +15q^4 _-}{96\pi q^3 _+ q^3 _- (q_+ +q_- )^5}
\end{equation}
where the two wave vectors $q_+ ,q_-$ verify $q^2 _+ q^2 _- =q^4 _*$ and $(q_+ +q_- )^2=\kappa^2+2q^2_*$.

The diagonal terms exhibit a divergence at high $q$ which can be removed by substracting the contribution of the neutral system as in section \ref{electrolyte}. The convergent term is given by:
\begin{equation}
\Delta G^{-1}_{c\,\sigma,\sigma}= \frac{24\pi l_B f^2}{a^2 c}\,I_2 (\mathbf{q})-\frac{1}{2}\left(\frac{48\pi l_B f^2}{a^2}\right)^2\,I_1 (\mathbf{q})
\end{equation}
with the new integral:
\begin{eqnarray}
\lefteqn{I_2 (\mathbf{q})=}\nonumber\\
& & \, \hspace{-1.5cm}\int\frac{d^3 \mathbf{q}_1}{(2\pi)^3} \frac{1}{\mathbf{q}^2 _1 (\mathbf{q}^2 -\mathbf{q}^2 _1)}\Bigg\{
\frac{\left( \mathbf{q}_1.(\mathbf{q}-\mathbf{q}_1)-\mathbf{q}^2 \right)^2}{(\mathbf{q}-\mathbf{q}_1)^2 ((\mathbf{q}-\mathbf{q}_1 )^2 +\kappa^2 )+q^4 _* }-\frac{(\mathbf{q}^2+\mathbf{q}^2_1 )(\mathbf{q}-\mathbf{q}_1)^2-(\mathbf{q}.\mathbf{q}_1)^2}{\mathbf{q}^2 _1 (\mathbf{q}^2 _1 +\kappa^2 )+q^4 _* } 
\Bigg\}\nonumber\\
\end{eqnarray}
The low $q$ expansion of this last integral is given by:
\begin{equation}
I_2 (q)=\frac{q^2}{6\pi q_+ q_- (q_+ +q_-)}
\end{equation}
\subsection*{A.2\ \ Exact expression for the salt-salt correction terms}
The diagonal and off-diagonal correction terms in the salt-salt matrix are equal after substraction of the unphysical divergence and the common value reads:
\begin{equation}
\Delta G^{-1}_{n\,\sigma,\sigma '}=-8\pi^2 l_B^2 \,I_3 (q)
\end{equation}
with the integral:
\begin{equation}
I_3 (q)=\int\frac{d^3 \mathbf{q}_1}{(2\pi)^3}\frac{\mathbf{q}^2 _1 (\mathbf{q} -\mathbf{q} _1)^2}{[\mathbf{q}^2 _1 (\mathbf{q}^2 _1 +\kappa^2 )+q^4 _* ][(\mathbf{q}-\mathbf{q}_1)^2 ((\mathbf{q}-\mathbf{q}_1 )^2 +\kappa^2 )+q^4 _* ]}
\end{equation}
Although we calculated exactly this integral we do not give here the result because it is lengthy and we do not need it for the discussion of the structure factor as it will appear clearly in the section \ref{discussion}.
\subsection*{A.3\ \ Exact expression for the salt-monomer correction terms} 
As in the salt-salt matrix, the diagonal and off-diagonal correction terms of the salt-monomer matrix are equal to:
\begin{equation}
\Delta G^{-1}_{c\,n\,\sigma,\sigma '}=-\frac{96(\pi l_B f)^2}{a^2} I_4 (q)
\end{equation}
with the following integral
\begin{equation}
I_4 (q)=\int\frac{d^3 \mathbf{q}_1}{(2\pi)^3}\frac{\left( \mathbf{q}^2-\mathbf{q}_1.(\mathbf{q}-\mathbf{q}_1) \right)}{[\mathbf{q}^2 _1 (\mathbf{q}^2 _1 +\kappa^2 )+q^4 _* ][(\mathbf{q}-\mathbf{q}_1)^2 ((\mathbf{q}-\mathbf{q}_1 )^2 +\kappa^2 )+q^4 _* ]}
\end{equation}
As in the preceding subsection, we will not give the precise evaluation of the integral for the sake of clarity.

\section*{Appendix B}
Figure 4 represents the $n-c$ translation of the $t-s$ phase diagram found in the main text. We have redefined the small ion concentration by substracting the polymer counterions; it is thus such that at $n=0$ there are still some small ions, the respective counterions of the polycations and polyanions. With this convention, the macroscopic spinodal line given by $st=1$ is written as:
\begin{equation}
\tilde{n}=\frac{l_B f^2}{\tilde{\chi}a}-f\tilde{c}
\end{equation}
with the following rescaled parameters:
\begin{eqnarray}
\tilde{n} & = & na^2l_B \\
\tilde{c} & = & ca^2l_B \\ 
\tilde{\chi} & = & \frac{\chi}{a^3}
\end{eqnarray}
The spinodal corresponding to microphase separation ($t=2-s$) is given by:
\begin{equation}
\tilde{n}=\frac{6^{1/2}f\tilde{c}^{1/2}}{\pi^{1/2}}-\left(f+\frac{3a\tilde{\chi}}{2\pi l_B}\right)\tilde{c}
\end{equation}
Finally the spinodal of complexation ($t=\frac{u}{(s+2)^{3/2}}$) is given by the formula:
\begin{equation}
\label{trinome}\tilde{n}=\frac{3l_B^{1/3}f^2}{2^{4/3}\pi^{2/3}\tilde{\chi}^{2/3}\tilde{c}^{1/3}}-\frac{2\,3^{1/2}f\,\tilde{c}^{1/2}}{(2\pi)^{1/2}}-f\tilde{c}
\end{equation}
At fixed $c,n$ and $\chi$, the critical value of the charge density for polyelectrolyte complexation is given by the solution of the quadratic Eq.\ref{trinome} in $f$.

\newpage
\section*{Figure Caption}
\begin{description}
\item[Figure 1] 1a.Typical phase diagram for a solution of oppositely charged polyelectrolytes at very low charge densities. In a wide range of this phase diagram the polymers undergo a polymer-polymer phase separation, the neutrality being ensured by the small ions. The thick lines are the tie lines indicating coexisting phases.\\
1b. Typical phase diagram for a solution of oppositely charged polyelectrolytes with higher charge densities. In a wide range of this phase diagram there is coexistence of dense phases containing both polymers and dilute phases with almost no polymers: there is thus complexation between the polyelectrolytes. The thick lines are the tie lines.
\item[Figure 2] Partial $t-s$ phase diagram: the different phases are indicated with their relative boundaries. This phase diagram can be associated to the representation used in Figure 1a: the polymers are either incompatible or compatible (soluble).
\item[Figure 3] Complete $t-s$ phase diagram: the effects of composition fluctuations are taken into account, resulting in the possibility of complexation for low $\chi$. When the charge density $f$ is increased, the other parameters being imposed, the system follows the double arrow sketched on the phase diagram, from the incompatibility region, through the compatibility region and finally through the complexation region. 
\item[Figure 4]
Complexation phase diagram in the concentration variables. The value of $\tilde{\chi}$ is fixed ($\tilde{\chi}=0.001$) and the complexation line is shown for various values of $f$ (from bottom to top $f=1/50,1/20,1/10$).
\end{description}

\newpage
\begin{figure}[p]
\includegraphics*[width=15cm]{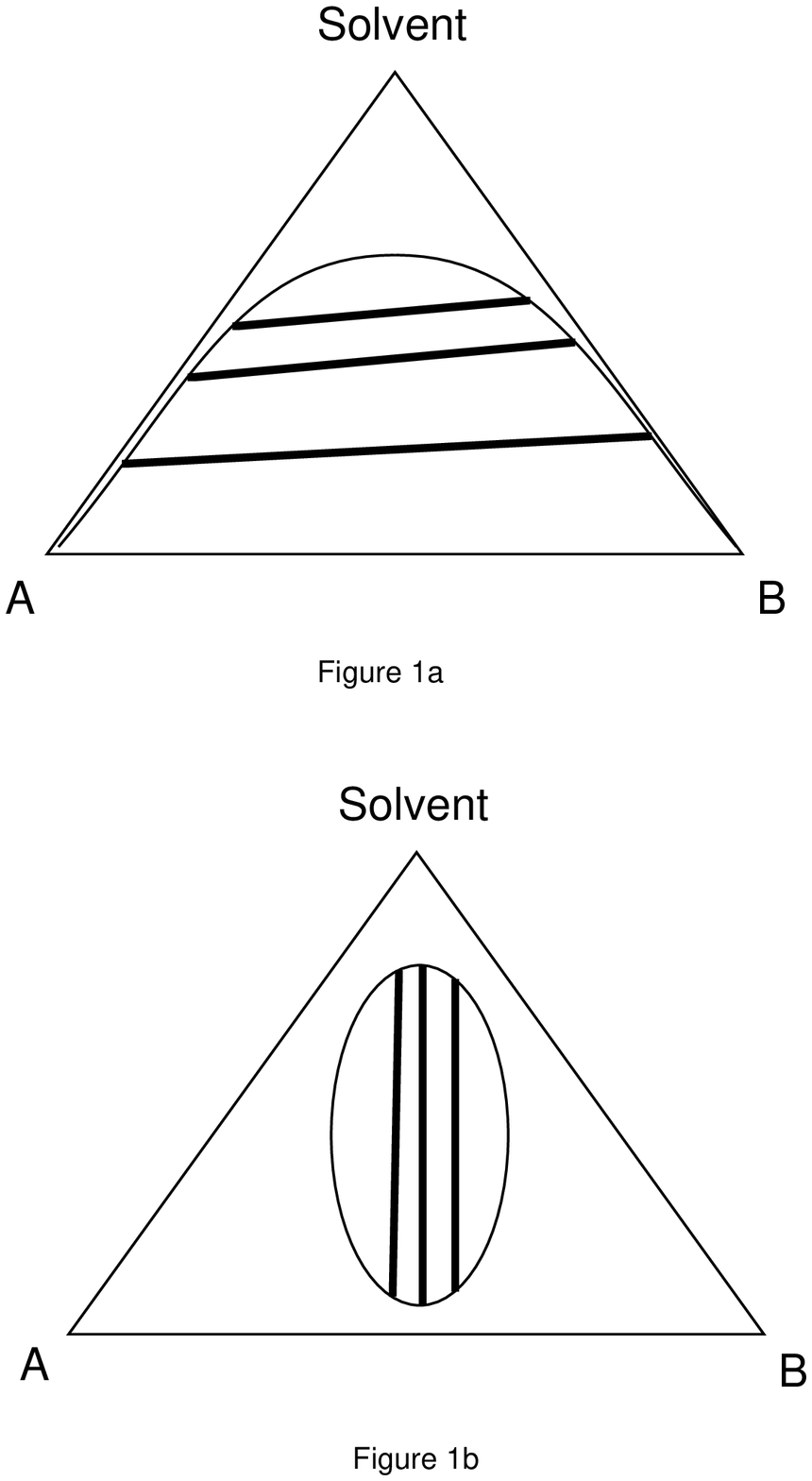}
\end{figure}
\newpage
\begin{figure}[p]
\includegraphics*[width=15cm]{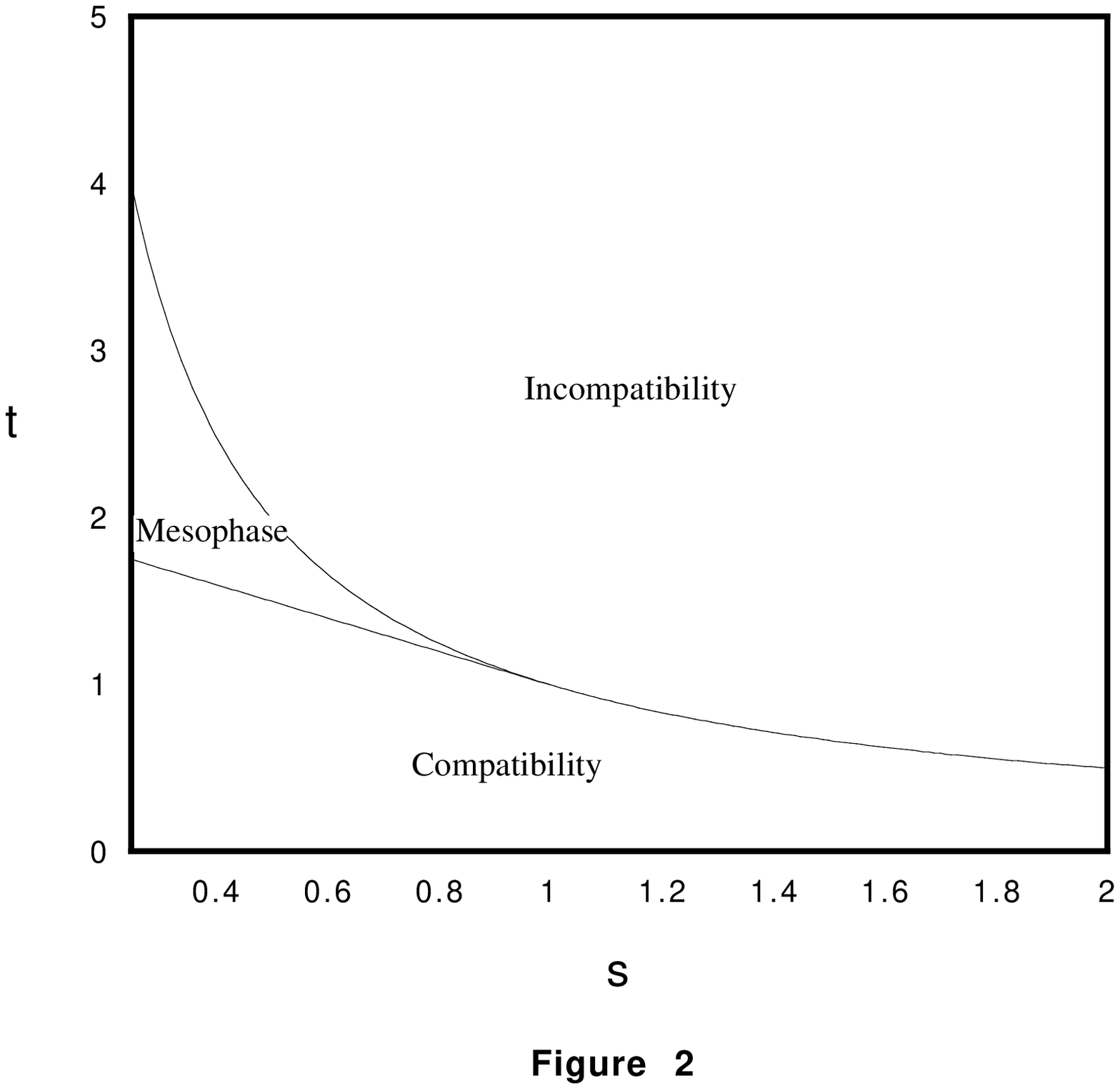}
\end{figure}
\newpage
\begin{figure}[p]
\includegraphics*[width=15cm]{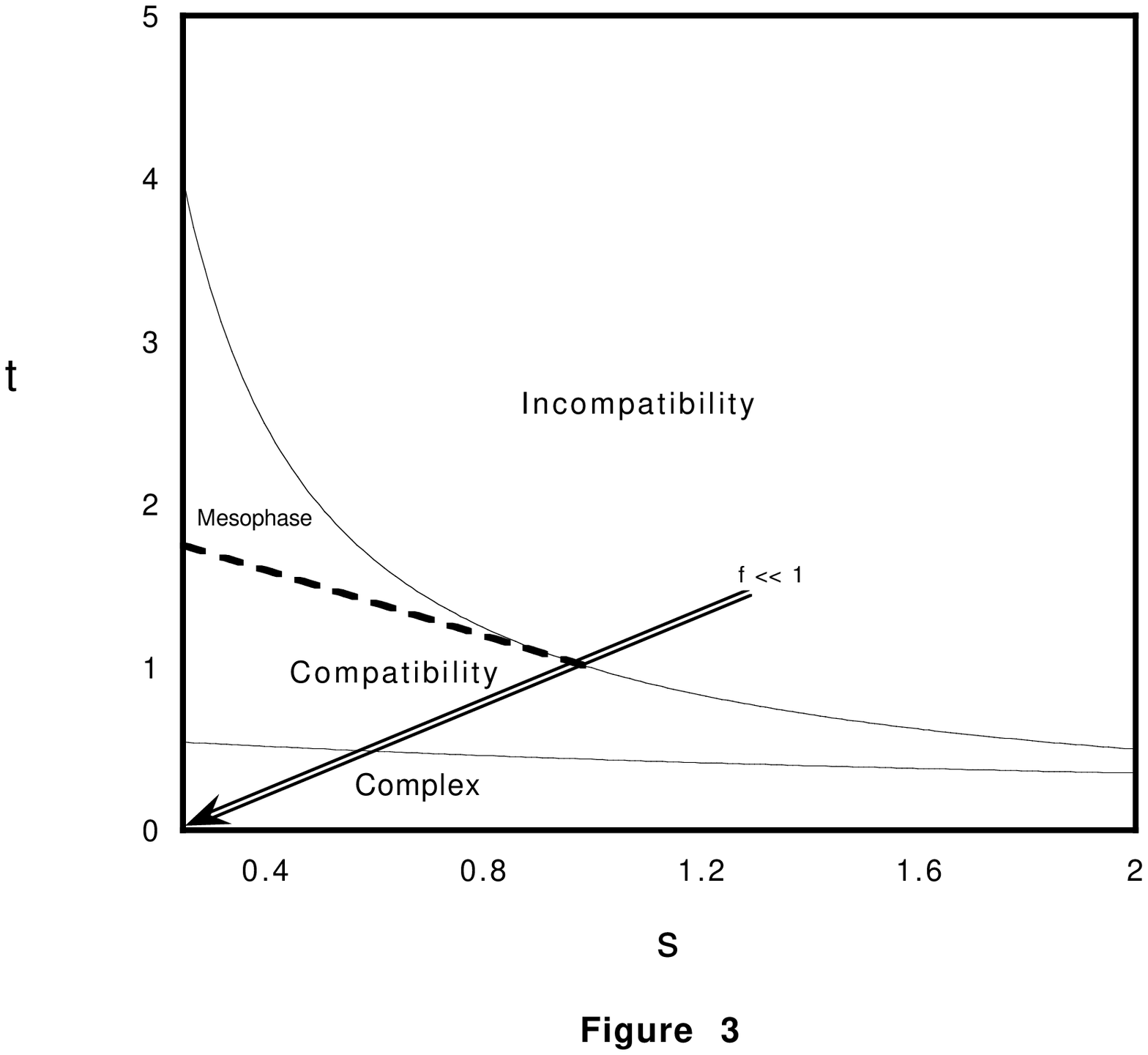}
\end{figure}
\newpage
\begin{figure}[p]
\includegraphics*[width=15cm]{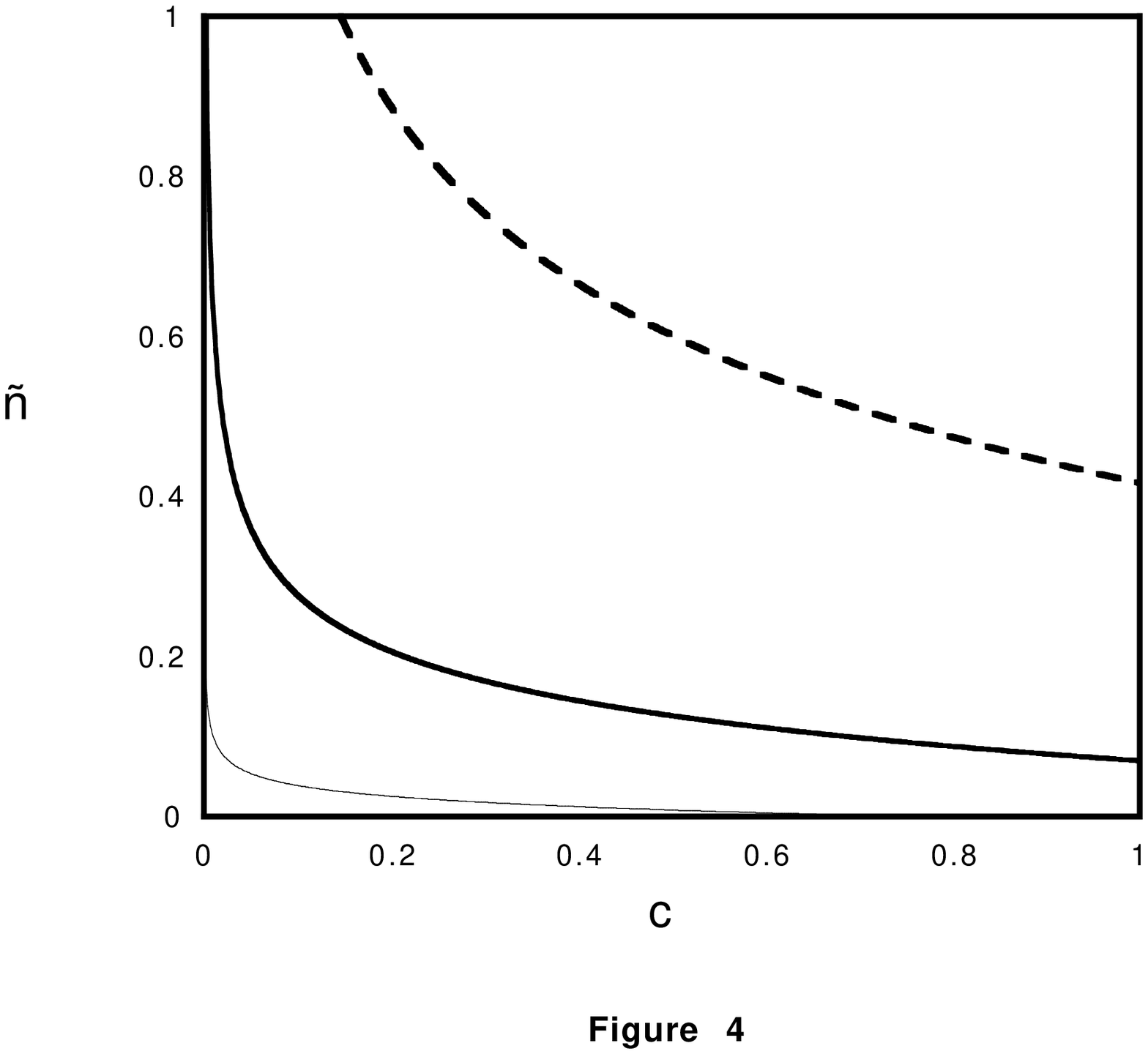}
\end{figure}

\end{document}